# Multi-Level Graph Attention Network Contrastive Learning for Knowledge-Aware Recommendation

Zhifei Hu[1] Feng Xia[2]
[1] Durham University, Durham, United Kingdom, knkw27@durham.ac.uk
[2] University of Sheffield, Sheffield, United Kingdom, fxia8@sheffield.ac.uk

**Abstract.** In recent years, the use of edge information provided by knowledge graphs along with the advantages of higher-order connectivity of graph networks in the context of recommendation systems has become a crucial research trend. However, this approach is commonly limited by the sparseness of labels, cannot properly learn the graph structure and a large number of noisy entities in the knowledge graph affects the precision of the recommendation results. To overcome this dilemma, herein, we propose a multi-view graph network contrastive learning approach. This method is capable of enhancing the user representation by employing the multi-view knowledge graph distillation technique to more accurately capture the user's preferences for relationships and entities. The network aggregates the neighborhood entity information of the item to create an appropriate representation of the item. A multi-level self-supervised comparative learning module is methodically established to perform multi-angle cross-comparison between internal features of users, between users and users, and between users and items from three levels: Inter-Level, Intra-Level, and Interaction-Level. Learning improves the ability to generalize the model to intraclass samples and the ability to discriminate inter-class samples and realize the multi-dimensional feature modeling. Herein, we conduct empirical evaluations through comparing the baseline and ablation models on three public datasets. The carried-out experiments reveal that the proposed model outperforms other current state-of-the-art baseline models, and the effectiveness of each model's module is also proved through ablation experiments.

## 1 Introduction

With the continuous spread and development of mobile phone networks, users receive daily a large amount of information, which is far beyond the ability of users to receive and process information and reduces the efficiency of users in information processing, the so-called information overload phenomenon. As a solution to this problem, the recommendation system is noticeably capable of reducing the information search time for users and has become a very crucial module in many Internet products. Early recommendation algorithms [1,2,3] follow the idea of collaborative filtering [4, 5, 6], which recommends items that are most similar to their hobbies and interests by evaluating the similarity between items and users. The collaborative filtering algorithm essentially utilizes the click information of users and items as the major modeling feature, and its fitability and generalizability are moderately limited.

In recent years, with the unceasing progress of knowledge graph technology [7], how to extract rich information in knowledge graphs to enhance the effect of a recommendation system has become a new research focus. The primary research work by combining a knowledge graph and recommendation system can be generally divided into two stages. First, knowledge representation learning is performed on the knowledge graph [8] and the representation vectors of entities and relationships are appropriately extracted, and then the two representation vectors are employed as downstream recommendation vectors. The input of the recommendation algorithm system based on the above method has certain defects. For instance, the representation vector of entities and relationships, which is learned by the representation learning algorithm for knowledge graphs, exhibits more of the structural information of the knowledge graph itself. The representation vector is more suitable for predicting missing entities and relations in the knowledge graph, but not quite suitable as input to the recommendation algorithm. In order to compensate for the shortcomings of the above methodologies and make better use of feature information in the knowledge graph for recommendation algorithms, researchers turned their attention to graph neural networks (GNNs) [9,10,11]. The knowledge graph recommendation algorithm based on the graph neural network [12, 13] exhibits excellent interpretability and scalability and can discover users' deep interests and hobbies. However, the proposed existing knowledge graph algorithms based on GNNs are often unable to perform sufficient learning due to the sparseness and unbalanced distribution of label data, and there are a large number of irrelevant noise entities in the knowledge graph, which will affect the final prediction and have a particular impact on the corresponding accuracy.

To address the above problems, introducing contrastive learning into graph neural networks recommender systems [14, 15, 16] has recently become an emerging field. Contrastive learning is a self-supervised representation learning based on the discriminative method [17, 18, 19]. The main idea of contrastive learning [20, 21] is to group similar samples and different samples that are far from each other, train the model, shorten the distance between positive samples, and reduce the distance between positive samples and negative samples. The major challenges of implementing contrastive learning in GNNs lie in how to design better pairs of positive and negative samples and how to choose better graph encoders for various datasets. Contrastive learning is capable of learning without additional label data and effectively reduces the cost of labeling. Moreover, compared to traditional classification problems, adaptive learning essentially learns the distance between positive and negative samples, and the influence of long-tailed data distribution is trivial. However, the proposed GNN-based contrastive learning technology has a particular perspective in designing positive and negative sample pairs and only employs the simple graph structure of the user-item bipartite graph, without taking complex knowledge graph data into account.

To solve the above problems, we design a multi-level knowledge graph contrastive learning (MGACL). First, we design a user information extraction module based on attention knowledge distillation to extract entity information from the perspective of entities and relationships. The process of calculating scores for entities from multiple perspectives helps create a more comprehensive user preference. Second, we design a graph-based convolution module to enhance the feature representation. Finally, we design a multi-level contrastive learning module that exploits full self-supervised learning to extract essential internal information from three levels, i.e., intra, inter, and interaction. The intra-level mainly starts from the features of users and items and considers the semantic features of the entity extracted from various perspectives as positive sample pairs. The inter-level employs noise to interfere with the representation of users and items and considers noisy entities and entities as positive samples. The interaction level is considered to be users and items that have interacted with each other in the presence of the same batchsize as positive samples. The negative samples of the three levels are all non-interacting user-item pairs in the same batch as negative samples. In view of the above declarations, our work contributions can be summarized as follows:

• Based on the idea of knowledge graph supervised learning for the recommendation, an unsupervised multi-level comparative learning approach is introduced, which realizes the mixed supervised and unsupervised learning paradigm. This reduces the problems concerning data noise and label sparsity during recommendation and enhances the recommendation accuracy.

• We propose a new MGACL recommendation model that uses multiple relational and entity perspectives to perform knowledge graph distillation on user preferences. It also incorporates a multi-level comparative learning task designed for user/item feature learning and provides a complete proof-theoretic analysis.

• We perform rich experiments on three publicly available datasets. The basic experimental results reveal that the proposed MGACL outperforms various advanced recommendation methods under different parameter settings. The achieved results by the ablation tests further indicate the rationality and effectiveness of each module of the model.

## 2 Related Work

### 2.1 Knowledge Graph in recommendation

In real recommendation scenarios, user behavior data is often very sparse, which readily leads to the overfitting of the model. The semantic relationship information between entities in the knowledge graph enriches the features of the items, reduces the common problem of cold start in recommender systems, helps to increase the variety of recommended items, increases the interpretability of the model, and improves robustness and generality of the model. The current knowledge graph-based recommendation technology includes three paradigms: embedding-based method, path-based method, and a combination of embedding and path. The constituent paradigms are briefly explained in the following:

**Embedding-based method:** This method shows the entities and relationships in the knowledge graph as dense vectors with low dimensions. Bordes et al. [31] proposed the TransE model on the basis of the translation idea, which assumed the relationship between entities in a knowledge graph as a kind of translation between entities. Wang et al. [32] conducted a deeper investigation based on the idea of the translation model, providing linkages between entities and complex relationships. Wang et al. [38] and Zhang et al. [39] introduced the knowledge representation learning technology into the recommender system and obtained the vector of entities and relations through the representation learning algorithm and used it as the peripheral information and entered it into the recommender system. In general, knowledge representation learning techniques focus more on changing the semantic relationships between entities and are often more suitable for node-type prediction and edge relationship prediction tasks in graphs than recommendation tasks.

**Path-based approach:** This method generates multiple source paths through a random walk with a manual or design algorithm and combines the vectors of dissimilar paths into one vector for a recommendation. Shi et al. [33] designed a meta-path-based random walk sampling methodology to generate multiple meta-paths starting from the target node and consisting of multiple nodes, and based on this, proposed a meta-path learning algorithm embedded in the HIN. Sun et al. [34] proposed the RKGE algorithm based on the recurrent network modeling, which compensates for the drawbacks of the embedding learning algorithm that only cares about the node semantic layer. Wang et al. [40] proposed a path-order-dependent algorithm. The KPRN, a relationship-based modeling algorithm, is capable of modeling the complex relationship of paths connecting users and items and distinguishing the importance of different paths through pooling operations. The use of resource path modeling to aggregate node neighborhood information highly relies on a large number of manual path designs and requires additional manual input when dealing with complex large-scale knowledge graphs.

**Embedding and path combination method:** In order to better integrate the knowledge graph into the recommender system, more and more works combine entity and relation representation learning with meta-path learning. In this regard, Wang et al. [7] proposed a KGAT algorithm based on a knowledge graph and attention mechanism, which signified graph relationships, users, and items in the same graph space, and employed a graph attention mechanism network to extract entities from mixed graphs with higher-order and semantics connectivity. Wang et al. [35] proposed a perceptual domain-based model KGCN, which searches for neighboring nodes in the knowledge graph according to the user's click history items, and utilizes the graph attention mechanism to unceasingly aggregate, update, and propagate. Wang et al. [36] proposed a water wave network, which cleverly borrowed the idea of water wave propagation and combined the representation-based learning method with the path-based methodology. The suggested approach was also capable of realizing the user's preference propagation, as well as exploring the user's potential interest and hobby.

### 2.2 Contrastive Learning

Contrastive learning [22,23] is a type of self-supervised learning whose main idea is to perform unsupervised learning on unlabeled data. As a result, the closer the distance between similar sample representations is, the farther the distance between various sample representations is to obtain the general characteristics of the dataset. Contrastive learning has been extensively implemented in the image, NLP, GNN, recommendation, and other fields. Zou D et al. [20] proposed a new framework of user/item representation learning, the so-called MCCLK, which learns user/item representations from two scopes—first, MCCLK is performed from the dimensions of global structural view, local synergistic view, and semantic view. Second, cross and multi-level comparative learning methodologies are commonly performed on these three views to arrive at more feature information. Yang et al. [24] designed a general knowledge graph contrastive learning (KGCL) framework to reduce the information noise problem of knowledge graph-augmented recommendation systems. Wu et al. [29] proposed an SGL framework to improve supervised recommendation tasks through self-supervised learning on user-item bipartite graphs. Ma X et al. [28] proposed a hybrid contrastive approach, namely graph-based recommendation learning that integrates unsupervised and supervised contrastive learning.

## 3 Theory and framework

We propose a multi-profile graph neural network contrastive learning recommendation model for multi-task learning based on a combination of user-item bipartite graphs and knowledge ones. The whole architecture of the model has been illustrated in Fig. 1, which essentially consists of two parts: graph convolution and self-supervised contrastive learning. The specific theoretical details of the model are provided in the following.

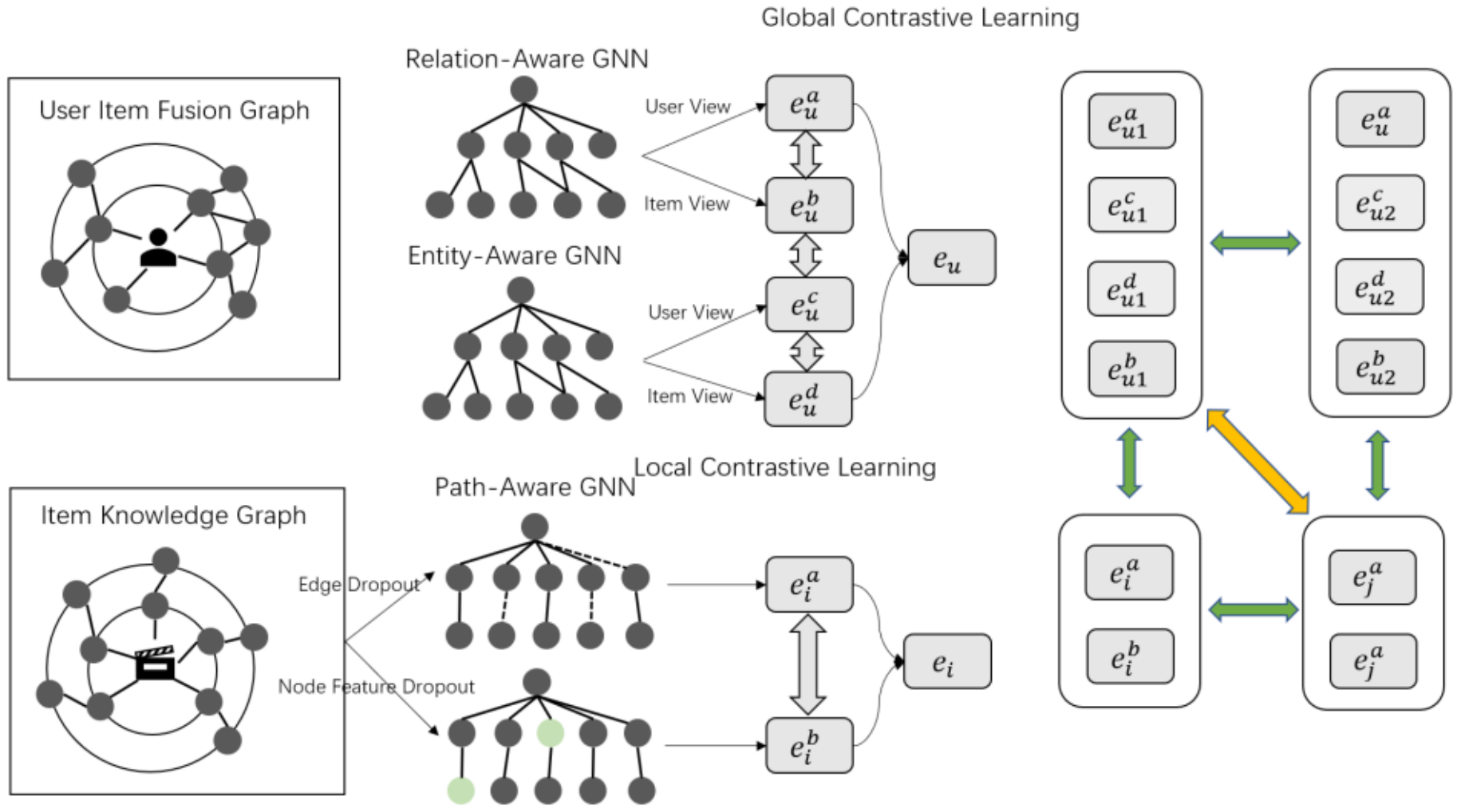

**Fig. 1.** The overall framework of the proposed MGACL. The framework is composed of three modules: Multi-view Knowledge Graph Distilling, GCN-Aware Item Representation and Multi-Level Contrastive Learning.

### 3.1 Task Definition

This section is aimed to introduce all the concepts and symbols used and appropriately gives the definition of the research task of the present work. In the recommendation scenario, we treat $u$ the interaction graph between users ($G_u$) and items as a bipartite graph ($v$). In order to enhance the accuracy of recommendation results, we introduce knowledge graphs ($G_k$). The definition of the bipartite graph ($G_u$) and the knowledge graph ($G_k$) are given as follows:

User and item bipartite graph ($G_u$): We take the user set $U = \{u_1, u_2 \ldots\}$ and the item set as $V = \{v_1, v_2 \ldots\}$ entities in the bipartite graph. If an edge connection exists between the user and the item $G_u$ in $G_u$, the interaction matrix $Y = \{y_{u,v} | u \in U, v \in V \ldots\}$ of $y_{u,v} = 1$ the user $u$ and the item $v$; otherwise, it implies that no edge connection exists between the points.

Knowledge graph ($G_k$): Knowledge graph $G_k = \{(h, r, t) | h, t \in \varepsilon, r \in R\}$, which $h$, r, $t$ signify the head entity, relationship, and tail entity of triples in the knowledge graph. $v$ represents any entity $N(v)$ in the knowledge graph, and is the set of neighbor entities of the entity in the knowledge graph $v$.

Entity set ($\varepsilon_u^p$): Based on the knowledge graph $G_k$ and the user-item bipartite $G_u$ graph, $G_H$ denotes the first $p$-hop entity set on $\varepsilon_u^p$. The items clicked by the user $G_k$ can be defined as follows:

$$\varepsilon_u^p = \left\{t \middle| \left(h, r, t \in G_H \; and \; h \in \varepsilon_u^{p-1}\right)\right\}, p = 1,2,3 \ldots l_p$$

In particular, in the case of $p$= 0, $\varepsilon_u^0 = \{v | y_{uv} = 1\}$, where $\varepsilon_u^0$ denotes the set of items clicked by the user in the history $u$.

Preference set ($S_u^p$): Define the user preference set $p$ of the first $\varepsilon_u^{p-1}$ hop as a triple set of head entities:$S_u^p$

$$S_u^p = \left\{(h, r, t) \middle| \left(h, r, t \in G_H \; and \; h \in \varepsilon_u^{p-1}\right)\right\},$$

$p = 1,2,3 \ldots l_p$

In particular, in the case of $p$= 0 , $S_u^0 = \{(u, \tilde{r}, \mathrm{t}) | t \in \varepsilon_u^0\}$, where $\tilde{r}$ represents the relationship of the user clicking on the item.

Research task: We define the research task as follows: given a user-item bipartite graph $G_u$ and a knowledge graph $G_k$, learning a prediction function $\mathrm{f} = (u, v; \Theta, G_u, G_k)$ that accurately predicts the possibility of a user $u$ clicking on an item $v$, where $\Theta$ denotes the parameter of the function $f$.

### 3.2 Multi -view knowledge graph distilling

Since different users have dissimilar preferences for relationships and entities in the hybrid graph $G_H$ , we evaluate the importance of neighborhood nodes from the perspectives of relationships and entities, resulting in better aggregating neighborhood information. This part chiefly introduces the theoretical approaches to generating user representation learning, including three parts: relation-based attention, entity-based attention, and user preference aggregation contrastive learning.

### 3.2.1 Relation -view Attention Mechanism

Given that users have different preferences for different relationships between entities in the knowledge graph. For example, when recommending the movie Avenger: Endgame, different users will have different choices for different relationships. Some users are interested in director Anthony Russo, while some users may prefer actor Chris Evans. We propose a relational perspective-based attention mechanism to enhance user representations. By calculating the correlation between the $\widetilde{\pi_{r_i}^u}$ user vector and the relationship in the preference set , the $S_u^p$ user's preference for the relationship is measured. The correlation score $\widetilde{\pi_{r_i}^u}$ calculation formula is as follows:

$$\pi_{r_i}^u = u^T r_i$$

$$\widetilde{\pi_{r_i}^u} = \frac{exp\ (\pi_{r_i}^u)}{\sum_{(h_i,\ r_i,\ t_i) \in S_u^p} exp\ (\pi_{r_i}^u)} \tag{1}$$

where $S_u^p$ is the preference set and $\widetilde{\pi_{r_i}^u}$ is $\pi_{r_i}^u$ the normalized score.

### 3.2.2 Entity-view Information Aggregation

Although evaluating the importance of neighbor entity nodes through relationships can characterize users' preferences for different entities, this approach is relatively simple and has great limitations. Because for the two entities with the same relationship, it is impossible to distinguish the importance of the entity through the relationship. To this end, we calculate the similarity of users, items, and entities in the preference set through the entity perspective to evaluate the importance of different entity nodes. Similar to Eq. (1), $\widetilde{\pi_{e_i}^u}$ and $\widetilde{\pi_{e_i}^v}$ the calculation formulas of user, item, and entity correlation scores are as follows:

$$\pi_{e_i}^u = u^T e_i$$

$$\widetilde{\pi_{e_i}^u} = \frac{exp\ (\pi_{e_i}^u)}{\sum_{(e_i h,\ r_i,\ e_i t) \in S_u^p} exp\ (\pi_{e_i}^u)} \tag{2}$$

$$\pi_{e_i}^v = v^T e_i$$

$$\widetilde{\pi_{e_i}^v} = \frac{exp\ (\pi_{e_i}^v)}{\sum_{(e_i h,\ r_i,\ e_i t) \in S_u^p} exp\ (\pi_{e_i}^v)} \tag{3}$$

### 3.2.3 User Preference Aggregation

Based on the relational perspective normalized correlation score $\widetilde{\pi_{r_i}^u}$ calculated by Eq. (1), we perform a weighted summation of the relational perspective correlation scores for the tail entities ($t_i$) of the preference set and aggregate to obtain the sum of all $S_u^p$ p-hop user preference responses from the relational perspective $O_u^r$:

$$O_r^p = \sum_{(e_i{}^h,\ r_i,\ e_i{}^t) \in S_u^p} \widetilde{\pi_{r_i}^u} t_i \ , p = 0,1,2,3 \ldots l_p \tag{4}$$

$$O_u^r = \sum_p O_r^p \ , \ p = 0,1,2,3 \ldots l_p \tag{5}$$

Based on the entity perspective normalized correlation score, which is calculated by the formula ( $\widetilde{\pi_{e_i}^u}, \widetilde{\pi_{e_i}^v}$) and Eq. (3), we perform the weighted summation of the entity perspective correlation score on the tail entities $t_i$ of the preference set $S_u^p$ and summarize all p-hop user preferences of the entity perspective sum of responses $O_e^p, O_u^e$:

$$O_e^p = \sum_{(e_i{}^h,\ r_i,\ e_i{}^t) \in S_u^p} (\widetilde{\pi_{e_i}^u} + \widetilde{\pi_{e_i}^v}) t_i \ , p = 0,1,2,3 \ldots l_p \tag{6}$$

$$O_u^e = \sum_p O_e^p \ , p = 0,1,2,3 \ldots l_p \tag{7}$$

Finally, the aggregate initializes $u$, the relational view as well as the sum of all p-hop user preference responses from the entity views $O_u^r$ and $O_u^e$ . The final vector representation of the user $O_u$ can be generated as:

$$O_u = u + O_u^r + O_u^e \tag{8}$$

The pseudocode for multi-view knowledge graph distillation is as follows:

---

Algorithm 1: Multi-view Knowledge Graph Distilling (MKGD)

---

**Input**: Fusion Graph $\boldsymbol{G_H}$

**Parameter**: $\{\boldsymbol{l_p}, \mathbf{M}\}$

**Output**: User representation $\boldsymbol{O_u}$

1: Let $\boldsymbol{O_u^r} = 0, \boldsymbol{O_u^e} = 0$.

2: **for** $p = 1, \ldots, l_p$ **do**

3: $\boldsymbol{O_r^p} \leftarrow \sum_{(e_i{}^h,\ r_i,\ e_i{}^t) \in S_u^p} \widetilde{\boldsymbol{\pi_{r_i}^u}} \boldsymbol{t_i}$ ;

4: $\boldsymbol{O_e^p} \leftarrow \sum_{(e_i{}^h,\ r_i,\ e_i{}^t) \in S_u^p} (\widetilde{\boldsymbol{\pi_{e_i}^u}} + \widetilde{\boldsymbol{\pi_{e_i}^v}}) \boldsymbol{t_i}$ ;

5: **end for**

6: **for** $p = 1, \ldots, l_p$ **do**

7: $\boldsymbol{O_u^r} \leftarrow \boldsymbol{O_u^r} + \boldsymbol{O_r^p}$;

8: $\boldsymbol{O_u^e} \leftarrow \boldsymbol{O_u^e} + \boldsymbol{O_e^p}$;

9: **end for**

10: **/* agg is sum aggregator function */**

11: $\boldsymbol{O_u} \leftarrow agg(u, \boldsymbol{O_u^r}, \boldsymbol{O_u^e})$**;**

12: **return** $\boldsymbol{O_u}$

---

### 3.3 GCN-Aware item representation

The knowledge graphs possess rich entity connection information. We design a graph convolutional network to extract entity-effective information from the knowledge graphs.

First, we implement the score function $\pi_{t_i}^{u}$ to calculate the score of user $u$ for neighborhood entities under $t$ relation $r$, which is exploited to represent the importance of neighborhood entities $t$ in representing user $u$ preference features. The calculation formula of $\pi_{t_i}^{u}$ is as follows:

$$\pi_{t_i}^{u} = w(u + r + t) + b \tag{9}$$

We normalize the attention score as follows:

$$\widetilde{\pi_{t_i}^{u}} = \frac{\boldsymbol{exp}\ (\pi_{t_i}^{u})}{\sum_{(h_i,\ r_i,\ t_i) \in \boldsymbol{s_u^p}} \boldsymbol{exp}\ (\pi_{t_i}^{u})} \tag{10}$$

The aggregation of the neighborhood information can be accomplished by using $\widetilde{\pi_{t_i}^{u}}$. The aggregated feature ($v_{N(v)}^{p}$) is calculated in the following form:

$$\boldsymbol{v}_{N(v)}^{p} = \sum \widetilde{\pi_{t_i}^{u}}\, e_{t_i} \tag{11}$$

Then, after obtaining the neighborhood information, it should be fused and updated with the central entity $v$. We use the summation approach to directly add the vectors to get the updated entity vector $v^{p}$:

$$v^{p} = \boldsymbol{v}_{N(v)}^{p} + v^{p} \tag{12}$$

Finally, the updated $v^{p}$ entity is evaluated as the neighborhood entity $v^{p-1}$of the first hop $p-1$, so that the information is propagated along the graph connection from the outside to the inside, and the neighborhood information $v_{N(v)}^{0}$ and central entity information of the zeroth layer are obtained as $v^{0}$. We use the final neighborhood information $v_{N(v)}^{0}$ and the central entity information $v^{0}$ as the positive samples of the item for comparative learning. As a result, the final item feature is evaluated as the sum of the neighborhood information and the central entity information.

### 3.4 Multi -Level contrastive learning

We use contrastive learning to increase the model robustness in recommendation tasks. The conventional contrastive learning approaches based on graph data augmentation only employ the user-item bipartite graphs and do not take the item knowledge graphs and mixed graphs into account. The common method of augmenting graph data by discarding graph nodes and edges is not capable of handling knowledge graphs with different types of entities and relationships properly. To remove this deficiency, we design positive and negative sample pairs of users and items based on the three levels of intra, inter, and interaction to extract the essential features of entities at the deep level.

### 3.4.1 Intra- Level

We believe that the semantics of a node could be determined by its neighboring nodes, so we take the item neighborhood information extracted by the graph convolution and the item semantic vector as pairs of positive samples within the item. Meanwhile, for users, we consider the features of the user entity perspective and the user relationship perspective as internal positive sample pairs of the user. For the sample comparison between users and items, we design the formulas $L_{_u_intra}$ and $L_{_v_intra}$:

$$L_{_u_intra} = \sum_{u \in U} -\log \frac{\exp(O_u^e \cdot O_u^r / \tau)}{\sum_{u,v \notin Y} \exp(e_u \cdot e_v / \tau)} \quad (13)$$

$$L_{_v_intra} = \sum_{v \in V} -\log \frac{\exp(O_v^s \cdot O_v^c / \tau)}{\sum_{u,v \notin Y} \exp(e_u \cdot e_v / \tau)} \quad (14)$$

Among them, $O_u^r$ and $O_u^e$ are, respectively, the aggregated information of user relationship perspective and user entity perspective, which are calculated by Eqs. (5) and (7), $O_v^s$ and $O_v^c$ in order represent the item neighborhood and the central entity features, $e_u$ and $e_v$ are the semantic vectors of users and items, respectively, and $\tau$ denotes the temperature coefficient. In general, the hierarchical contrastive learning loss function ($L_{intra}$) is defined by:

$$L_{_intra} = \frac{L_{_u_intra} + L_{_v_intra}}{2} \quad (15)$$

### 3.4.2 Inter- Level

In order to enhance the differences between users and items, we randomly discard the user representations generated in Section 3.2 ( $O_u$ ) and the item representations generated in Section 3.3 ($O_v$) with a certain probability and get the $\widehat{O_u}$ sum $\widehat{O_v}$. We regard $O_u$ sum $\widehat{O_u}$ as a user-side positive sample pair, and $O_v$ and $\widehat{O_v}$ as an item-side positive sample pair. Comparative learning is able to effectively improve the model's ability to learn items and user features, and enhance the robustness of the model. We devised the formula $L_{_u_inter}$ and $L_{_v_inter}$:

$$L_{_u_inter} = \sum_{u \in U} -\log \frac{\exp(\boldsymbol{O_u} \cdot \widehat{\boldsymbol{O_u}} / \tau)}{\sum_{u,v \notin Y} \exp(e_u \cdot e_v / \tau)} \quad (16)$$

$$L_{_v_inter} = \sum_{v \in V} -\log \frac{\exp(\boldsymbol{O}_v \cdot \widehat{\boldsymbol{O}}_v / \tau)}{\sum_{u,v \notin Y} \exp(e_u \cdot e_v / \tau)} \tag{17}$$

where $e_u$ and $e_v$ represent the initialization semantic vectors of users and items, respectively, and $\tau$ stands for the temperature coefficient. Under the same batch, the items with which the user experiences no interaction are $e_v$

$$L_{_inter} = \frac{L_{_u_inter} + L_{_v_inter}}{2} \tag{18}$$

### 3.4.3 Interaction-Level

At the interaction level, we take the interacting user and item as a positive pair. In the presence of the same batch, the items with which the user has no interaction are regarded as negative sample pairs at the interaction level ($O_{v'}$). The formula for calculating the contrastive learning loss at the interaction-level ($L_{interaction}$) is provided as follows:

$$L_{_interaction} = \sum_{u \in U, v \in V} -\log \frac{\exp(\boldsymbol{O}_{\boldsymbol{u}} \cdot \boldsymbol{O}_v / \tau)}{\sum_{<u,v \in >Y, <u,v'> \notin Y} \exp\left(\boldsymbol{O}_{\boldsymbol{u}} \cdot \boldsymbol{O}_{v'} / \tau\right)} \tag{19}$$

## 3.5 Multi-task training

We train contrastive and recommender learning systems together, both having the same parameter space. For the recommendation task, we utilize the cross-entropy loss function to evaluate the recommendation loss ($L_{base}$) as follows:

$$L_{_base} = -\sum_{y_{uv} \in N} [y_{uv} \log(\hat{y}_{uv}) + (1 - y_{uv}) \log(1 - \hat{y}_{uv})] \tag{20}$$

We adopt a joint learning approach for training the model, and the total loss function of the model ($L_{total}$) can be calculated as follows:

$$L_{_total} = L_{_base} + \lambda_1 (L_{interaction} + L_{intra} + L_{inter}) + \lambda_2 ||\mathcal{F}||_2^2 \tag{21}$$

where $\lambda_1$ denotes the coefficient of the comparative learning loss, $||\mathcal{F}||_2^2$ denotes the 12 regular terms of the model parameters, and $\lambda_2$ represents the coefficient of the regular term.

# 4 Experiments

In this section, we evaluate the proposed MGACL model on three real-world datasets and verify the effectiveness of the model. Our goal is to answer the following questions:

- RQ1: How does the MGACL perform compared to those state-of-the-art baseline models?
- RQ2: How do various components affect the MGACL?
- RQ3: How do various settings of parameters influence the MGACL?

## 4.1 Experimental Settings

### 4.1.1 Dataset and hyper-parameter settings

To evaluate the effectiveness of the model, we use three public recommendation datasets: Yelp-2018[7], Last-fm[35], and Amazon-book[30] for the sake of empirical analysis. First, the dataset scores are binarized, and then 20-core is applied to filter the dataset, filtering the user behavior records with fewer than 20 interactions. Finally, the entities are matched against the freebase open-source knowledge graph to construct their corresponding sub-knowledge graphs. A detailed description of the dataset and hyperparameter settings has been presented in Table 1.

**Table 1.** Dataset & Hyper-parameter Settings

| Stats. | | Yelp-2018 | LFM-1b | AZ-book |
|---|---|---|---|---|
| User-Item Interaction Network | #Users | 45919 | 12134 | 6969 |
| | #Items | 45538 | 15471 | 9854 |
| | #Interactions | 1183610 | 2979267 | 552706 |
| | #Avg user clicks | 25.77 | 152.3 | 79.3 |
| | #Avg clicked items | 25.99 | 119.4 | 56.1 |
| Knowledge Graph | #Source | | Freebase | |
| | #Entities | 136499 | 106389 | 113487 |
| | #Relations | 42 | 9 | 39 |
| | #Triples | 1853704 | 464567 | 2557746 |
| Hyper-parameter Settings | $\boldsymbol{l_p}$ | 2 | 3 | 3 |
| | $\boldsymbol{l_h}$ | 2 | 2 | 3 |
| | $\boldsymbol{M}$ | 64 | 32 | 64 |
| | $\boldsymbol{N}$ | 3 2 | 8 | 16 |
| | $\boldsymbol{t}$ | 0.2 _ | 0.2 _ | 0.2 |

We further present the optimal hyperparameter settings on the three datasets. As demonstrated in Table 1, for Yelp-2018, LFM-1b, and AZ-book, user sampling depth $l_p$, item sampling depth $l_h$, user neighborhood sampling number $M$, item neighborhood sampling number $N$, and temperature coefficient $t$ are set as [2, 3, 3], [2,2,3], [64,32,64], [32,8,16], and [0.2,0.2,0.2], respectively.

### 4.1.2 Baseline Methods

To verify the effectiveness of the MGACL, we compare it with the following baseline models on the above three datasets, which are:

(1) **NFM [27]** is a specific approach that combines a neural network with factorization, specifically, using the second-order cross-term of the factorization as input to the neural network model to extract deep semantic features.
(2) **CKE [37]** combines multimodal information with recommendation tasks and employs graph representation learning to extract structural and semantic features of entities in knowledge graphs, which is aimed to provide implicit information for recommendation tasks.
(3) **BPRMF [28]** constructs clicked/unclicked items triples to train user-item vectors, which is capable of properly learning under highly sparse data conditions.
(4) **RippleNet [36]** utilizes the user's historical clicked items to display the user's feature representation, each clicked item corresponds to an entity in the knowledge graph, and the information surrounding the entity in the knowledge graph is appropriately extracted for modeling.
(5) **KGCN [35]** maps the current predicted item to the knowledge graph, finds the surrounding information of the item in the knowledge graph, collects the information from outside to inside through the graph convolution algorithm, and finally forms the item representation.
(6) **KGAT [7]** is an end-to-end framework that combines a bipartite graph of users and items with a knowledge graph and uses graph convolution to collect the information of entities in the knowledge graph and improve the representation of items and users.
(7) **SGL [29]** combines contrastive learning with a graph encoder to augment the graph data by randomly removing, and masking the features of nodes and edges to construct positive and negative samples, so that the model learns more basic features.
(8) **MVIN [30]** learns items' representations from multiple user and entity perspectives, enriches user-item interactions, and refines entity-entity interactions.

**Table 2.** AUC、ACC and F1 Results on all datasets

| Models | Yelp-2018 | | LFM-1b | | Amazon-book | |
|---|---|---|---|---|---|---|
| | AUC | ACC | AUC | ACC | AUC | ACC |
| NFM | 0.919(-3.47%) | 0.851(-3.95%) | 0.946(-2.87%) | 0.879(-4.46%) | 0.842(-6.55%) | 0.764(-6.60%) |
| CKE | 0.922(-3.15%) | 0.854(-3.61%) | 0.948(-2.67%) | 0.881(-4.24%) | 0.836(-7.21%) | 0.745(-8.92%) |
| BPRMF | 0.919(-3.47%) | 0.852(-3.84%) | 0.946(-2.87%) | 0.878(-4.57%) | 0.832(-7.66%) | 0.742(-9.29%) |
| RippleNet | 0.928(-2.52%) | 0.858(-3.16%) | 0.938(-3.70%) | 0.887(-3.59%) | 0.820(-8.99%) | 0.745(-8.92%) |
| KGCN | 0.915(-3.89%) | 0.844(-4.74%) | 0.917(-5.85%) | 0.865(-5.98%) | 0.808(-10.3%) | 0.733(-10.4%) |
| KGAT | 0.916(-3.78%) | 0.846(-4.51%) | 0.937(-3.80%) | 0.875(-4.89%) | 0.852(-5.44%) | 0.780(-4.65%) |
| SGL | 0.916(-3.78%) | 0.851(-3.95%) | 0.948 (-2.67%) | 0.878 (-4.57%) | 0.891 (-1.11%) | 0.799(-2.32%) |
| MVIN | 0.941 (-1.16%) | 0.869 (-1.92%) | 0.965 (-0.92%) | 0.910 (-1.09%) | 0.875(-2.89%) | 0.793 (-3.06%) |
| MGACL | 0.952 ( %) | 0.886 ( %) | 0.974 ( %) | 0.920 ( %) | 0.900 ( %) | 0.818 ( %) |

**Table 3.** Top@N NDCG, RECALL Results on all datasets and all experiments are performed 3 times.

| Models | Yelp-2018 | | LFM-1b | | Amazon-book | |
|---|---|---|---|---|---|---|
| | NDCG | RECALL | NDCG | RECALL | NDCG | RECALL |
| NFM | 0.2139 | 0.0211 | 0.2371 | 0.0119 | 0.1913 | 0.0395 |
| CKE | 0.2124 | 0.0211 | 0.2375 | 0.0115 | 0.2442 | 0.0529 |
| BPRMF | 0.2270 | 0.0223 | 0.2369 | 0.0126 | 0.2409 | 0.0536 |
| RippleNet | 0.1101 | 0.0129 | 0.2173 | 0.0115 | 0.1208 | 0.0151 |
| KGCN | 0.2089 | 0.0081 | 0.1585 | 0.0079 | 0.1717 | 0.0341 |
| KGAT | 0.2015 | 0.0102 | 0.2062 | 0.0102 | 0.2628 | 0.0692 |
| SGL | 0.0383 | 0.0113 | 0.1857 | 0.0454 | 0.1017 | 0.0561 |
| MVIN | 0.0675 | 0.0055 | 0.1244 | 0.0139 | 0.2421 | 0.0539 |
| MGACL | 0.3104 | 0.0221 | 0.7866 | 0.0744 | 0.3492 | 0.0437 |

## 4.2 Performance Comparison (RQ1)

We compared MGACL with other baseline models in three datasets and selected AUC, ACC, and F1 as evaluation indicators to assess the predictive ability of the model. The experimental results are given in Table 2. In addition, we selected RECALL, NDCG to evaluate the Top@20 model results, and the experimental results are presented in Table 3. The experimental results indicate that our model achieves the best results on all datasets compared to the state-of-the-art baseline models. Specifically:

• As can be seen from Table 2, the proposed model MGACL has achieved the highest indicators in Yelp-2018, LFM-1b, and Amazon-book datasets. Compared to the baseline models that obtained the best results in each data set, MGACL exhibits the highest performance, revealing 1.16%, 0.92%, and 1.11% improvement with respect to the above datasets. In general, the advantages of MGACL are mainly reflected in the following: 1) Knowledge graph-based contrastive learning is capable of reducing the influence of irrelevant entities in the graph on the construction of user and item features as well as extracting the basic characteristics of entities. 2) Contrastive learning based on the interaction level increases the negative sample size of the user-item and enhances the generalization ability of the model in dealing with different entities.

• In particular, we compare the performance of MGACL on different datasets. MGACL achieves the largest improvement on the Amazon Book, while it performs moderately on the LFM-1b. This is mainly ascribed to the fact that the Amazon Book data is more scattered compared to the LFM-1b, and the MGACL can effectively improve the prediction accuracy and benefit from the effective signals provided by self-supervised tasks.

• As shown in Table 3, among all NDCG@20 metrics, our MGACL model improves by 36%, 50%, and 42% compared to the best baseline model in the LFM-1b, Amazon-book, and yelp-2018 datasets, respectively. This is essentially attributed to the fact that we introduce contrastive learning compared to the MVIN model, and compared to the SGL, the knowledge graph has been included.

• Among all baseline models, MVIN outperforms other models on Yelp-2018 and Last fm, while SGL performs best on Amazon-book. On the LFM-1b and Yelp-2018 datasets, the CF-based method is relatively better than the KG-based method, and the KG-based method is better than the CF method in the Amazon Book. This is because the Amazon book has relatively little interaction information between users and items, so the CF method cannot learn more effectively. In the three datasets, the CL method performs best in Amazon book and worst in the Yelp-2018, which may be due to the specificity between different datasets, for instance, the sparsity of datasets.

• We compare the knowledge graph-based models of the CKE, RippleNet, KGAT, KGCN, and MVIN. It is found that the MVIN performs the best on the three datasets, while the KGCN performs the worst on the three datasets. This is because MVIN personalizes GNN from the user's perspective and introduces a deep and extensive structure to perform multi-view enhanced representation learning for users/items. However, the KGCN only takes the neighborhood entities of the items in the knowledge graph as input and does not integrate the bipartite graph of the user-item with the item KG.

• As shown in Table 2, the performance of the baseline models is ranked in three datasets LFM-1b, Yelp -2018, and Amazon-book. This is usually caused by the difference in the number of user-item interactions in the three datasets. As a general rule, the more interactions, the more thoroughly the model learns.

### 4.3 Comparison subcomponents (RQ2)

We conduct ablation tests to demonstrate the effectiveness of each module. The proposed model is studied from the perspective of user interaction, entity perspective, item graph complexity, and contrastive learning, and the following four types of the MGACL are designed for experiments:

- w/o RV: remove the user relationship view module.
- w/o EV: eliminate the user entity view module.
- w/o GCN: remove the item graph complexity module.
- w/o CL: removes the multi-level contrastive learning module.

**Table 4.** MGACL ablation study results.

| Ablation | Components | | | | Datasets | | |
|---|---|---|---|---|---|---|---|
| | RV | EV | GCN | CL | LFM | AZ | Yelp |
| | | | | | AUC | AUC | AUC |
| N/A | ✔ | ✔ | ✔ | ✔ | 0.9739 | 0.9004 | 0.9523 |
| w/o RV | | ✔ | ✔ | ✔ | 0.9735 | 0.9000 | 0.9517 |
| w/o EV | ✔ | | ✔ | ✔ | 0.9620 | 0.8990 | 0.9520 |
| w/o GCN | ✔ | ✔ | | ✔ | 0.9721 | 0.8964 | 0.9500 |
| w/o CL | ✔ | ✔ | ✔ | | 0.9642 | 0.8750 | 0.9484 |

The ablation experimental results are presented in Table 4 and the following results can be concluded:

• All four modules have a specific contribution to the final test results. The introduction of contrastive learning is the most advanced in the model. Even if there is a lack of information aggregation from an institutional or relational perspective, it can learn well. The second is the graph convolution module.

• In contrast, using the user entity perspective module alone is superior to using the user relationship perspective module alone. For the Yelp-2018 dataset, the relational perspective module performs slightly better than the entity perspective module. This may be ascribed to the discrepancies in knowledge graphs. The greater the number of relationships, the greater the role of the user relationship perspective module.

• Among the three datasets, contrastive learning performed best on the Amazon book dataset, followed by the LFM and Yelp. This may be attributed to the fact that different datasets exhibit various sensitivities to the construction of positive and negative sample pairs.

## 4.4 Parameter Sensitivity (RQ3)

In this part, we explore the performance of the MGACL under different parameters, as clarified in some detail as follows:

**Table 5.** User-item Sampling depth.

| | AZ-book | | Yelp-2018 | | LFM-1b | |
|---|---|---|---|---|---|---|
| | AUC | F1 | AUC | F1 | AUC | F1 |
| $l_p = 1$ | 0.8924 | 0.8119 | 0.9482 | 0.8792 | 0.9633 | 0.9326 |
| $l_p = 2$ | 0.8919 | 0.8135 | **0.9488** | **0.8742** | **0.9668** | **0.9377** |
| $l_p = 3$ | **0.8930** | **0.8121** | 0.9486 | 0.8812 | 0.9682 | 0.9383 |
| $l_p = 4$ | 0.8929 | 0.8108 | 0.9477 | 0.8805 | 0.9679 | 0.9387 |
| $l_h = 1$ | 0.8930 | 0.8121 | 0.9488 | 0.8742 | 0.9682 | 0.9383 |
| $l_h = 2$ | 0.8939 | 0.8165 | **0.9500** | **0.8871** | **0.9694** | **0.9401** |
| $l_h = 3$ | **0.8954** | **0.8150** | 0.9491 | 0.8844 | 0.9673 | 0.9377 |
| $l_h = 4$ | 0.8943 | 0.8099 | 0.9498 | 0.8761 | 0.9674 | 0.9364 |

• User-item sampling depth $l_p$&$l_h$: We set $l_p$and $l_h$in the range of {1, 2, 3, 4}, and perform experimental comparison studies on three datasets, including AZ-book, Yelp-2018, and LFM-1b. As presented in Table 5, the MGACL achieves the best results for these datasets in the cases of $l_p = \{3, 2, 3\}$ and $l_h = \{3, 2, 2\}$, respectively. For the dataset AZ-book, MGACL adopts a larger user-item sampling depth, which may be because AZ-book exhibits less user-item interaction information, and a larger sampling depth is required to mine the graph data deeper, thereby assisting the recommendation task.

**Table 6.** User&Item Neighborhood Sampling Size.

| | AZ-book | | Yelp-2018 | | LFM-1b | |
|---|---|---|---|---|---|---|
| | AUC | F1 | AUC | F1 | AUC | F1 |
| $M = 8$ | 0.8954 | 0.8150 | 0.9500 | 0.8871 | 0.9694 | 0.9401 |
| $M = 16$ | 0.8963 | 0.8133 | 0.9505 | 0.8781 | 0.9722 | 0.9431 |
| $M = 32$ | 0.8957 | 0.8065 | 0.9496 | 0.8714 | **0.9738** | **0.9414** |
| $M = 64$ | **0.8978** | **0.8172** | **0.9510** | **0.8766** | 0.9718 | 0.9413 |
| $N = 8$ | 0.8978 | 0.8172 | 0.9510 | 0.8766 | **0.9738** | **0.9414** |
| $N = 16$ | **0.9004** | **0.8172** | 0.9512 | 0.8683 | 0.9731 | 0.9433 |
| $N = 32$ | 0.8944 | 0.7988 | **0.9523** | **0.8847** | 0.9736 | 0.9443 |
| $N = 64$ | 0.8886 | 0.7843 | 0.9501 | 0.8683 | 0.9733 | 0.9441 |

• We set the number of user and item neighborhood sampling size M&N: M and N are set in the range of {8, 16, 32, 64} and the experimental comparison studies are performed on three datasets, i.e., AZ-books, Yelp-2018, and LFM. As presented in Table 6, it is found that the best effect of the MGACL is achievable for these three datasets in the cases of M={64, 64, 32}, N={16, 32, 8}, respectively.

The numbers of user neighborhood samples are often larger than those of items. This may be ascribed to the fact that the user's interests are more scattered in the entity, whereas the item information is relatively concentrated, and the neighborhood samples of more items introduce noisy entities.

**Table 7.** Tempture.

| | AZ-book | | Yelp-2018 | | LFM-1b | |
|---|---|---|---|---|---|---|
| | AUC | F1 | AUC | F1 | AUC | F1 |
| $t = 0.1$ | 0.8990 | 0.8104 | 0.9478 | 0.8708 | 0.9739 | 0.9396 |
| $t = 0.2$ | **0.9004** | **0.8172** | **0.9523** | **0.8847** | **0.9739** | **0.9414** |
| $t = 0.3$ | 0.8592 | 0.7851 | 0.9466 | 0.8757 | 0.9485 | 0.9226 |
| $t = 0.4$ | 0.8639 | 0.7793 | 0.9473 | 0.8828 | 0.9417 | 0.9177 |
| $t = 0.5$ | 0.8589 | 0.7833 | 0.9452 | 0.8782 | 0.9478 | 0.9233 |
| $t = 0.6$ | 0.8652 | 0.7928 | 0.9484 | 0.8862 | 0.9448 | 0.9197 |
| $t = 0.7$ | 0.8654 | 0.7947 | 0.9447 | 0.8718 | 0.9432 | 0.9163 |

Temperature coefficient $t$: We tried all values of temperature coefficients $t$ between 0.1 and 0.7 on the three datasets as presented in Table 7, and to our surprise, almost all three datasets achieve the best performance in the case of $t = 0.2$.

## 5 Conclusion and future work

In this work, we investigate the limitations of knowledge graph-based recommendation combined with graph neural networks under the supervised learning paradigm and explore ways to introduce contrastive learning into knowledge graph recommendation. In particular, we propose an MGACL multi-level graph attention contrastive learning framework that alleviates the problem of noisy entities in knowledge graphs and the problem of label sparsity in supervised learning. The user representation is enhanced by distilling multi-perspective knowledge of relationships and entities on users. The convolutional graph network is employed to mine the neighborhood information of multiple items in the KG and complete the side information. Three distinct levels of semantic contrastive learning modules are designed to assist the task of supervised recommendation. We conducted several experiments on public datasets and demonstrated the superiority of the MGACL. This work provides a new exploration for the research on the contrastive learning in knowledge graph recommendation and opens up novel research opportunities. In future work, we hope to try various approaches for constructing positive and negative sample pairs for contrastive learning, such as generating sample pairs using improved graph structure to provide more supervised signals.